\documentclass[fleqn,usenatbib,useAMS]{mnras}

\makeatletter
\gdef\@journal{\@microfiche \hfill \@printed}
\def\@oddfoot{}
\def\@evenfoot{}
\makeatother

\usepackage[utf8]{inputenc}
\usepackage[T1]{fontenc}
\usepackage{ae,aecompl}

\usepackage{amsmath}  
\usepackage{txfonts}  

\DeclareUnicodeCharacter{B1}{\ensuremath{\pm}}

\title{Uncertainty propagation with functionally correlated quantities}

\author[M. Giordano]{Mosè Giordano\thanks{E-mail: mose.giordano@le.infn.it}\\
  Dipartimento di Matematica e Fisica `\emph{E. De Giorgi}',
  Università del Salento, Via per Arnesano, CP 193, I-73100 Lecce, Italy \\
  INFN, Sezione di Lecce, Via per Arnesano, CP 193, I-73100 Lecce, Italy}

\pubyear{2016}

\begin{document}
\label{firstpage}
\pagerange{\pageref{firstpage}--\pageref{lastpage}}
\maketitle

\begin{abstract}
  Many uncertainty propagation software exist, written in different programming
  languages, but not all of them are able to handle functional correlation
  between quantities.  In this paper we review one strategy to deal with
  uncertainty propagation of quantities that are functionally correlated, and
  introduce a new software offering this feature: the Julia package
  \texttt{Measurements.jl}.  It supports real and complex numbers with
  uncertainty, arbitrary-precision calculations, mathematical and linear algebra
  operations with matrices and arrays.
\end{abstract}

\begin{keywords}
  methods: data analysis -- methods: numerical
\end{keywords}



\section{Introduction}
\label{sec:introduction}

\subsection{Measurement uncertainty}
\label{sec:meas-uncert}

The process of measurement has as goal the determination of the value of the
\emph{measurand}, that is the physical quantity to be measured.  The measurement
of physical quantities with a continuous spectrum is always hindered by
uncertainty.  The JCGM's ``Guide to the expression of uncertainty in
measurement''~\citep{gum} lists some of the possible sources of uncertainty in a
measurement.  These include experimenter's bias in reading analogue instruments,
finite instrument resolution, inexact values of constants or measurement
standards, approximations and assumptions in the measurement procedure, etc.
Thus, the outcome of any measurement must be specified giving a \emph{nominal
  value} and a \emph{measurement uncertainty}.\footnote{An exception is
  represented by measurements of those discrete measurands that can be simply
  counted, so their true value can be assessed with certainty.}

The uncertainty is the best estimate, according to the experimenter, of the
distance between the experimental result of a measurement and the true value of
the measurand up to a certain probability.  In this sense, the uncertainty
identifies the interval in which the experimenter is confident that the true
value of the measurand lies.

\subsection{Functional correlation between quantities}
\label{sec:funct-corr-betw}

The fact that two or more quantities are correlated means that there is some
sort of relationship beetween them.  In the context of measurements and error
propagation theory, the term ``correlation'' is very broad and can indicate
different things.  Among others, there may be some dependence between
uncertainties of different measurements with different values, or a dependence
between the values of two measurements while their uncertainties are different.
Thus, it is very important to specify what kind of correlation one is talking
about.

Here, by ``functional correlation'' we mean the functional relationship between
quantities: if \(x = \bar{x} \pm \sigma_x\) is an independent measurement, a
quantity \(y = f(x) = \bar{y} \pm \sigma_y\) that is function of \(x\) is not
like an independent measurement, but is a quantity depending on \(x\), so we say
that \emph{\(y\) is functionally correlated with \(x\)}.

Scientists analysing the result of experiments usually have to perform
mathematical operations involving measurements, whose uncertainty must be
appropriately propagated to the outcome of the operations.  This is very
important, considering that many quantities cannot be measured directly, but one
has to measure another measurand and transform its value to the desired quantity
through mathematical operations.

When functional correlation is taken care of, one should find
\(x - x = 0 \pm 0\), \(x/x = 1 \pm 0\), because \(x\) is perfectly correlated
with itself and there is no doubt that the difference between a measurement and
itself is exactly \(0\).  Likewise, its own ratio is exactly \(1\).  The same
applies to any other expression, like \(x^{2} = xx\),
\(\tan(x) = \sin(x)/\cos(x)\),
\(\mathrm{cis}(x) - \exp(\mathrm{i}x) = 0 \pm 0\), \dots.  If functional
correlation is not taken into account, you would find that the two sides of each
equations would have different uncertainties.

In Section~\ref{sec:uncer-prop} we review one approach to deal with uncertainty
propagation of operations involving functionally correlated quantities.  In
Section~\ref{sec:measurements-jl} we present one software that uses this
strategy: the Julia package \texttt{Measurements.jl}.

\section{Uncertainty propagation and functional correlation}
\label{sec:uncer-prop}

For a function \(G(a, b, c, \dots)\) of real arguments with uncertainties
(\(a = \bar{a} \pm \sigma_{a}\), \(b = \bar{b} \pm \sigma_{b}\) and
\(c = \bar{c} \pm \sigma_{c}\), \dots) the linear error propagation theory
prescribes that uncertainty is propagated as
follows~\citep{taylor1997introduction}:
\begin{equation}
  \begin{split}
    \sigma_G^2 &= \left( \left.\frac{\partial G}{\partial a}\right\vert_{a =
        \bar{a}} \sigma_a \right)^2 + \left( \left.\frac{\partial G}{\partial
          b}\right\vert_{b = \bar{b}} \sigma_b \right)^2 + \left(
      \left.\frac{\partial G}{\partial c}\right\vert_{c = \bar{c}} \sigma_c
    \right)^2 + \cdots \\
    &\quad{}+ 2 \left(\frac{\partial G}{\partial a}\right)_{a = \bar{a}}
    \left(\frac{\partial G}{\partial b}\right)_{b = \bar{b}} \sigma_{ab} + 2
    \left(\frac{\partial G}{\partial a}\right)_{a = \bar{a}}
    \left(\frac{\partial G}{\partial c}\right)_{c = \bar{c}}
    \sigma_{ac} \\
    &\quad{}+ 2 \left(\frac{\partial G}{\partial b}\right)_{b = \bar{b}}
    \left(\frac{\partial G}{\partial c}\right)_{c = \bar{c}} \sigma_{bc} +
    \cdots,
  \end{split}
\end{equation}
where the \(\sigma_{ab}\) factors are the covariances defined as
\begin{equation}
  \sigma_{ab} = \mathrm{E}[(a - \mathrm{E}[a])(b - \mathrm{E}[b])].
\end{equation}
\(\mathrm{E}[a]\) is the expected value of \(a\).  If uncertainties of the
quantities \(a\), \(b\), \(c\), \dots, are independent and normally distributed,
the covariances are null and the above formula for uncertainty propagation
simplifies to
\begin{equation}
  \sigma_G^2 = \left( \left.\frac{\partial G}{\partial a}\right\vert_{a
      = \bar{a}} \sigma_a \right)^2 + \left( \left.\frac{\partial
        G}{\partial b}\right\vert_{b = \bar{b}} \sigma_b \right)^2 + \left(
    \left.\frac{\partial G}{\partial c}\right\vert_{c = \bar{c}} \sigma_c
  \right)^2 + \cdots.
\end{equation}

In general, calculating covariance terms is not an easy task.  One possible
approach for handling functional correlation is to propagate the uncertainty
always using really independent variables, so that covariances are null by
definition.  Thus, dealing with functional correlation boils down to finding the
set of all the independent measurements on which an expression depends and
calculating its partial derivatives with respect to all these quantities.

Going back to the example above, if \(a\), \(b\), \(c\), \dots, are correlated
quantities, while \(\{x, y, z, \dots\}\) is the set of really independent
measurements, it is possible to calculate the uncertainty of
\(G(a, b, c, \dots)\) with
\begin{equation}
  \label{eq:unc-prop-no-cov}
  \sigma_G^2 = \left( \left.\frac{\partial G}{\partial x}\right\vert_{x
      = \bar{x}} \sigma_x \right)^2 + \left( \left.\frac{\partial
        G}{\partial y}\right\vert_{y = \bar{y}} \sigma_y \right)^2 + \left(
    \left.\frac{\partial G}{\partial z}\right\vert_{z = \bar{z}} \sigma_z
  \right)^2 + \cdots
\end{equation}
where all covariances due to functional correlation are null.  If other types of
correlation (not functional) between \(x\), \(y\), \(z\), \dots, are present,
they should be treated by calculating the covariances as shown above.

For a function of one argument only, \(G = G(a)\), there is no problem of
correlation and the uncertainty propagation formula in the linear approximation
simply reads
\begin{equation}
  \sigma_G = \left\vert \frac{\partial G}{\partial a} \right\vert_{a =
    \bar{a}} \sigma_a
\end{equation}
even if \(a\) is not an independent variable and comes from operations on really
independent measurements.

As a concrete example, suppose you want to calculate the function
\(G = G(a, b)\) of two arguments \(a\) and \(b\) that are functionally
correlated, because they come from some mathematical operations on really
independent measurements \(x\), \(y\), \(z\), say \(a = a(x, y)\),
\(b = b(x, z)\).  By using the chain rule, the uncertainty on \(G(a, b)\) is
calculated as follows:
\begin{equation}
  \begin{split}
    \sigma_G^2 &= \left( \left(\frac{\partial G}{\partial a}\frac{\partial
          a}{\partial x} + \frac{\partial G}{\partial b}\frac{\partial
          b}{\partial x}\right)_{x = \bar{x}} \sigma_x \right)^2 + \Biggl(
      \left(\frac{\partial G}{\partial a}\frac{\partial a}{\partial y}\right)_{y
        = \bar{y}} \sigma_y \Biggr)^2 \\
    &\quad{}+ \left( \left(\frac{\partial G}{\partial
          b}\frac{\partial b}{\partial z}\right)_{z = \bar{z}} \sigma_z
    \right)^2.
  \end{split}
\end{equation}

\section{Measurements.jl}
\label{sec:measurements-jl}

I wrote and developed \texttt{Measurements.jl},\footnote{It is developed at
  \url{https://github.com/giordano/Measurements.jl}, where it is also possible
  to report issues and suggest improvements.}  a package that allows users to
define numbers with uncertainties, perform calculations involving them, and
easily get the uncertainty of the result according to linear error propagation
theory. This library is written in Julia, a modern high-level, high-performance
dynamic programming language designed for technical
computing~\citep{2012arXiv1209.5145B}.  \texttt{Measurements.jl} is free and
open source and is released under the terms of MIT ``Expat'' license.

The main features of the package are:
\begin{itemize}
\item Support for most mathematical operations available in Julia standard
  library, including special functions, involving real and complex numbers.  All
  existing functions that accept \texttt{AbstractFloat} (and
  \texttt{Complex\{AbstractFloat\}} as well) arguments and internally use
  already supported functions can in turn perform calculations involving numbers
  with uncertainties without being redefined.  This greatly enhances the power
  of the package without effort for the users.
\item Functional correlation between variables is correctly handled.
\item Support for arbitrary precision (also called multiple precision) numbers
  with uncertainties.  This is useful for measurements with very low relative
  error.
\item Define arrays of measurements and perform calculations with them.  Some
  linear algebra functions work out-of-the-box, including solution of linear
  systems, matrix multiplication and dot product between vectors, calculation of
  inverse, determinant, and trace of a matrix, QR decomposition.
\item Propagate uncertainty for any function of real arguments (including
  functions based on C/Fortran calls), using \texttt{@uncertain} macro.
\item Functions to get the derivative and the gradient of an expression with
  respect to one or more independent measurements.
\item Functions to calculate standard score and weighted mean.
\item Parse strings to create measurement objects.
\item Easy way to attach the uncertainty to a number using the \(\pm\) sign as
  infix operator.  This makes the code more readable and visually appealing.
\item Combined with external packages allows for error propagation of
  measurements with their physical units.
\end{itemize}

When used in the Julia interactive session, it can serve also as an easy-to-use
calculator.

Furthermore, \texttt{Measurements.jl} strives to be as fast as possible.  To
give a rough idea of its speed, according to tests performed with the
\texttt{BenchmarkTools.jl}
suite\footnote{\url{https://github.com/JuliaCI/BenchmarkTools.jl}.} on a system
equipped with an Intel(R) Core(TM) i7-4700MQ CPU, defining a number with
uncertainty takes \(30\)~ns, and summing two numbers with uncertainty requires
about \(400\)~ns.

\subsection{Handling functional correlation}
\label{sec:handl-funct-corr}

The package defines a new data type, \texttt{Measurement}, which holds the
nominal value of the measurement and its uncertainty, assumed to be a standard
deviation.  In order to deal with functional correlation between measurements
when performing mathematical operations with arity larger than one, a
\texttt{Measurement} object keeps inside the list of its derivatives with
respect to the independent variables from which the quantity comes in the form
of a dictionary.

When the type constructor is used to create a \texttt{Measurement} object, a
\emph{new independent} measurement is defined.  Instead, the outcome of any
mathematical operation involving \texttt{Measurement} objects is a quantity that
depends on all the quantities it comes from.  From a technical standpoint, it is
a \texttt{Measurement} object, not tagged as independent, which holds the list
of derivatives with the respect to the really independent quantities.  In this
way, the strategy presented in Section~\ref{sec:uncer-prop} can be readily
applied.

\subsection{Why a new uncertainty propagation package?}
\label{sec:why-new-package}

Among the packages listed in the Wikipedia article ``List of uncertainty
propagation software''~\citep{wiki:list-unc-prop-software}, to date
\texttt{Measurements.jl} is one of the most advanced and feature-rich.  It is
one of the very few programs supporting complex measurements and capable of
performing linear algebra operations with matrices and arrays of numbers with
uncertainties.  In addition, it is the only program that can work with arbitrary
precision arithmetic.  All operations are always carried out taking care of
functional correlation between quantities.

\texttt{Measurements.jl} is not the first uncertainty propagation software
implementing the method reviewed in Section~\ref{sec:uncer-prop}.  Actually, it
borrowed the idea of keeping the list of derivatives from the Python package
\texttt{uncertainties},\footnote{\url{https://pypi.python.org/pypi/uncertainties/}.}
but the rest of the implementation of \texttt{Measurements.jl} is completely
independent from that of \texttt{uncertainties}.  However, writing an
uncertainty propagation software in Julia language has some advantages.

The language itself is specifically designed for scientific computing with
particular attention to performance, approaching that of statically-compiled
languages like C and Fortran, and it has been being adopted by more and more
researchers across the world, so it was natural to make available to
experimental scientists such a tool purely implemented in Julia.

In addition, Julia language has a smart type system that greatly improves
productivity and lets users focus on the real problem at hand.
\texttt{Measurement} type is defined as a subtype of \texttt{AbstractFloat} type
and inherits all features of the parent type, thus support for complex
measurements, arbitrary precision calculations, array operations and linear
algebra in \texttt{Measurements.jl} came for free during the development of the
package, there is not a single line in the whole code of the program to reach
these features.  Also the possibility of combining \texttt{Measurements.jl} with
third-party packages to define numbers with uncertainty and physical units is a
feature that came without specific effort from the authors of the different
packages, thanks to the powerful Julia type system.  This is an important factor
in terms of maintainability of the code and productivity.

\subsection{Examples}
\label{sec:examples}

Here is a showcase of some examples of use of the \texttt{Measurements.jl}
package.

The code below shows how to define numbers with uncertainties and perform
operations with them.
\begin{verbatim}
using Measurements # Load the package

l = 0.936 ± 1e-3; T = 1.942 ± 4e-3
g = 4pi^2*l/T^2
# => 9.797993213510699 ± 0.041697817535336676
\end{verbatim}

This second example show that the functional correlation between quantities is
correctly handled, within numerical accuracy.
\begin{verbatim}
x = 8.4 ± 0.7 # An independent measurement
u = 2x # This is functionally correlated with x
(x + x) - u
# => 0.0 ± 0.0
u/2x
# => 1.0 ± 0.0
u^3 - (2x^3 + 6x*x^2)
# => 0.0 ± 0.0
cos(x)^2 - (1 + cos(u))/2
# => 0.0 ± 0.0
beta(x, u) - gamma(x)*gamma(u)/gamma(3x)
# => 0.0 ± 1.852884572118782e-23
\end{verbatim}

Other comprehensive examples presenting all the features of the package can be
found in the up-to-date documentation at
\url{http://measurementsjl.readthedocs.io}.

\section{Conclusions}
\label{sec:conclusions}

In Section~\ref{sec:uncer-prop} we reviewed a method to handle uncertainty
propagation in operations involving functionally correlated quantities.  The
expedient proposed entails tracking the true independent measurements from which
an expression comes and computing its partial derivatives with respect to those
measurements.  In this way the covariance terms are null by definition and the
simple equation~\eqref{eq:unc-prop-no-cov} can be used to propagate the
uncertainty.

This method has been implemented in the Julia package \texttt{Measurements.jl},
presented in Section~\ref{sec:measurements-jl}.  This software enables
scientists to perform fast operations on measured quantities while correctly
propagating their uncertainty to the result, according to linear error
propagation theory.  \texttt{Measurements.jl} features support for real and
complex numbers with uncertainty, multiple precision arithmetic, mathematical
and linear algebra operations with matrices and arrays of numbers with
uncertainty.

\section*{Acknowledgements}
\label{sec:acknowledgements}

I would like to thank Steven G. Johnson, for his valuable suggestions about the
design of \texttt{Measurements.jl}, and Marta Dell'Atti, for her useful comments
about this preprint.


\bibliographystyle{mnras}
\bibliography{bibliography}

\label{lastpage}
\end{document}